# Device to Remotely Track and Locate the Position of a Child for Safety

S.M.K.C.S.B. Egodawela, H.M.D.M.B. Herath, R.D.Ranaweera and J.V. Wijayakulasooriya

shamendrae@gmail.com,dushan.herath@eng.pdn.ac.lk,rdbranaweera@ee.pdn.ac.lk,jan@ee.pdn.ac.lk

*Abstract* - **Parents are always worried about the wellbeing of their children. As per the Statistics Report 2017 by Missing Children Europe Organization, a child is reported missing every 2 minutes. Due to the imminent threat, parents are prone to buy their children mobile phones to keep in touch with them. However, giving a Mobile phone to a child can cause issues including cyber bullying, improper use of social networks, access to mature age and illicit content on the internet and possibly, phone theft. As an effort to tackle some of those issues, this paper proposes a solution which enables parents to call, locate and track their children using a child-friendly mobile device. The common scenario the device would come to play is in enhancing the safety of a child who would travel alone on a typical route; for instance a child who walks from home to school and back. The device can be calibrated to keep track of a typical route of travel. Then, if the device detects some deviation from the usual route, it would trigger a notification to parents. A probability matrix based novel algorithm is introduced to detect route deviation. Design details of the mobile device, along with the details of the route deviation detection algorithm are presented in this paper.**

## INTRODUCTION

According to Missing Children Europe Organization report[1], 50,000 children are reported missing every year in the EU; that would be 1 child per every 2 minutes. Further, records by the U.S. Department of Justice Office of Juvenile Justice and Delinquency Prevention [2] indicate that children of ages 5 to 16 are most likely to be subjected to such abuse. Due to the imminent threat, parents are prone to buy their children Mobile phones to be in touch with them. However, there are some well know socio-psychological issues associated with handing a Mobile phone to a child, such as cyber bullying, improper use of Social Networking Media, access to inappropriate content on the Internet, and possibly phone theft.

Children and early adolescents tend to seek a degree of independence from parents as part of their growing up, and sometimes tend to do their day-to-day travel unaccompanied by a parent or a guardian. The modern working culture may also limit the opportunity for parents/guardians to accompany children on their daily travel. These have raised concerns regarding the safety of children, paving way to technologies and devices which enable child locating and tracking [3-8], also considerations towards ethics should be respected in relation to locating and tracking people [8].

In this paper, we propose a child-friendly mobile device equipped with a novel probability-based tracking algorithm. The typical use of the device will involve the device being carried or worn by children during their day-to-day travel—when going to school for example—and enabling parents to locate and track children, while facilitating parent-child communication (i.e., calling) as well. Among the current devices available in the market for such tracking purposes [9],[10], it was identified that most devices provide merely a GPS coordinate boundary for a route of travel a child may be taking [3-8]. This approach has merits in its simplicity and low cost of data processing. As a complement to such approaches, in this paper we propose a novel algorithm to detect route deviation. Our algorithm utilizes a basic probabilistic framework to send an alarm message to warn the parents in case the child's mobile device detects a route that is highly unlikely, with respect to previously observed routes. Here, rather than setting a boundary limit for the possible GPS coordinates data of previously observed routes, as done in [9][10], the previously observed routes are used to calculate the probability of the device being carried along a particular route. This probability is compared with a threshold and if the probability is less than the threshold, the corresponding route is deemed unusual or highly unlikely, and a route deviation alarm is sent to a smart device carried by a parent or guardian through an application. An advantage of our probabilistic method is the reduction of false alarms caused by the device taking a route slightly different to the normal, which was identified as a major drawback of contemporary devices in the market [9],[10].Additionally, in case of an emergency, e.g., in a kidnapping situation, the device enables the parent/guardian to remotely turn on the microphone of the device carried by the child to hear the developments in the vicinity. In order to stop any abuse of the device by the child, the child's mobile device will only be able to place calls to a single phone number, that of the parent.

This paper provides a comprehensive report of the design specifications, device parameters, and justifies the components and their selection. Further, an exhaustive explanation is provided of the proposed route deviation detection algorithm. The remainder of this paper is organized as follows: Functions of the device Section depicts the functionality of the device and explains the hardware used. Methodology Section justifies the selection of the microcontroller and explains its specifications and how the embedded system is constructed using each of the components together with how they function simultaneously. Further, it provides the specification and justification for selecting each component. Route deviation Detection Section provides the route deviation detection algorithm. The Result Section discusses the major difficulties that were faced in fabrication and in interfacing the hardware. This paper is concluded in Conclusions Section and the future work is proposed.

## FUNCTIONS OF THE DEVICE

In this section the functionality of the device will be presented. It was identified that the first step of synthesizing an embedded system is preparing the Functional design of the system. The basic functionality is as shown in Figure I and the additional features are produced henceforth in this section.

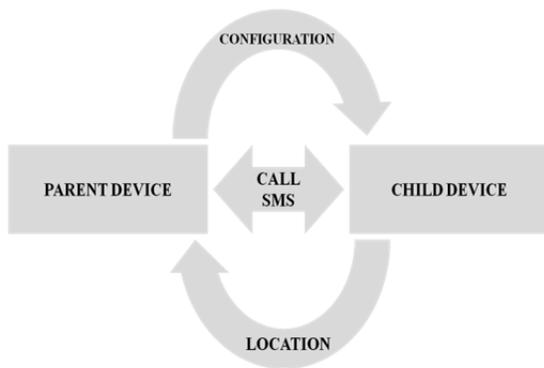

FIGURE I

The basic functions of the embedded system. The device can communicate only with the parent device. Moreover, the configuration of the device can only be done by an authorized smart device and cannot be accessed by any other. A location request when interrupted by the parent device will return the GPS data to track and locate the child device.

### A. Get GPS location (Global positioning system)

The Global Positioning System (GPS) is a satellite-based localization system made up of at least 24 satellites at an altitude of 20,000 km. Each satellite transmits a unique signal which includes information of position, current time at regular intervals and orbital parameters that allow GPS devices to decode and compute the precise location of the satellite. GPS receivers use this information and a method of trilateration to calculate the user's exact location.

The device was equipped with a GPS receiver to pinpoint the exact location of the device for locating and tracking purposes.

### B. Track live GPS location

When the parent device calls for live location tracking, the device sends a request for GPS localization at a periodic interval of five seconds. The location data is then sent to the parent device via a GSM module.

### C. GSM (Global System for Mobile Communications) service

GSM is a digital mobile standard that uses a variation of TDMA (Time Division Multiple Access). GSM digitizes and compresses voice data and channels with two other streams of user data, each in its own time slot. It operates at either the 900, 1800 or 1900 MHz frequency bands. Due to the ease of switching between carrier devices, GSM network is the most widely used communication technology worldwide.

A GSM compatible module SIM800L® was used in the device to communicate between the parent device and the child device. The criteria for the selection of this GSM module will be discussed later in Section IV. The module facilitates voice over communication and messaging between pre-authorized GSM carriers. The authorization could be provided only by the parent device to eliminate any communication between other GSM carriers.

### D. Ability to remotely turn on the microphone and speaker

The aim of the device is to remotely monitor the child device using the parent device. It was proposed to integrate the ability to turn on the in-built microphone. It will enable the parent device to listen to the developments in the vicinity. Thus, the ability to remotely turn on the speaker and microphone was facilitated in the design.

### E. SOS button in case of an emergency

It was suggested that in case of an emergency an easy access button/switch was placed on the device. The triggering of the button will place a call to the parent's mobile and send an SOS message by interrupting the GSM module. The SMS message will contain the last known GPS tag of the child device

### F. Low battery warning

Once the device reaches a lower threshold value of charge left in the battery it will send a *low battery warning* SMS. The message will contain the last known GPS tag of the child device.

After the functional design was done the implementation in hardware was carried out as described in Section III.

## METHODOLOGY

The embedded system was designed with the flow of functionality as shown in Figure II. Selection criteria for the rest of the hardware will be discussed here.

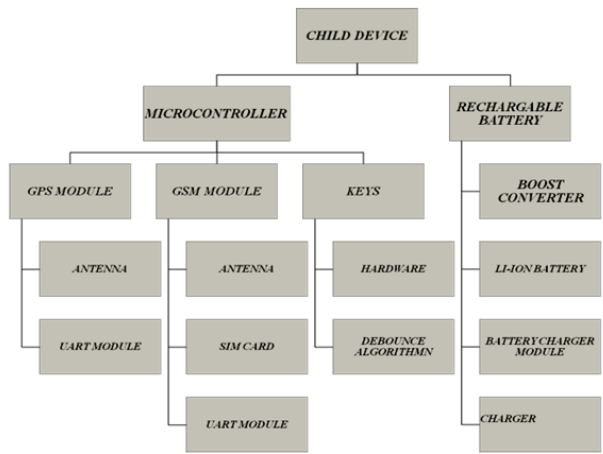

FIGURE II

The functional design of the system. The sub-processes and components as shown in figure were identified for the ease of design and implementation.

As shown in Figure III the proposed embedded system comprises of different protocols to communicate between components.

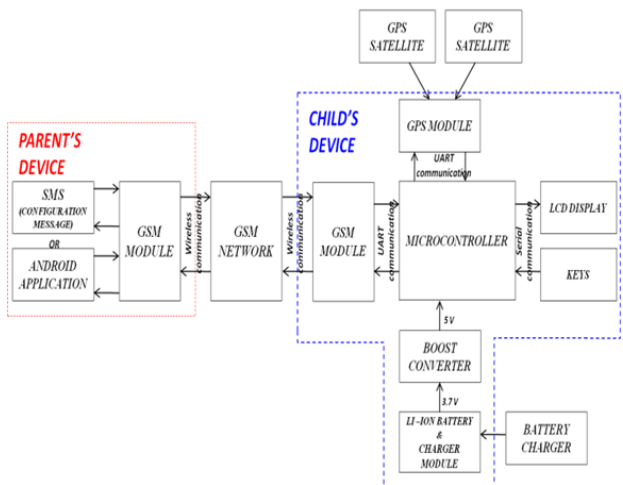

FIGURE III

The macro-level functional block diagram of the mobile device. The figure shows the components and their interface protocols. Additionally, it shows how the device operates in coordination with Global services such as GPS and GSM networks.

The selection of hardware after the detailed design is presented here.

*G. Atmega2560 Microcontroller*

After a survey of the functions and the modules that were to be used, the microcontroller was chosen considering the following criteria.

- 2 UART(Universal Asynchronous Receiver Transmitter) ports for communication between GPS and GSM modules.
- 4-IO Pins to control the LCD screen.
- 4-IO Pins for buttons
- MCLR reset
- About 16 MHz clock speed
- **Flash memory > 100 KB**
- **SRAM > 4KB**
- **EEPROM > 1KB**
- Working voltage 5V
- General purpose pin for SLEEP operation

A preliminary market survey was done of the available controllers. The ATmel ATmega2560® with Advanced RISC Architecture satisfied all the above requirements.

*H. SIM800L GSM Module*

The GSM module was chosen such that the device can be used with any mobile network in Sri Lanka. GSM 900 and GSM 1800 are the common 2G bands used in Sri Lanka. The SIM 800L® module which operates at frequency bands GSM 850MHz, EGSM 900MHz, DCS 1800MHz and PCS 1900MHz [12]. The module has merits in having a SLEEP mode and having low power consumption. However the working voltage of 3.4-4.4 V which is less than 5 V was a drawback. Further, it has speaker output and microphone input. The module is relatively small in size hence, beneficial for the overall weight and sizing of the device. Figure IV. shows the schematic of the connections made in hardware.

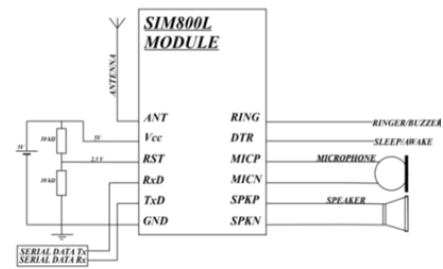

FIGURE IV

The SIM800L module hardware schematic. UART communication protocol facilitates the data transmission between microcontroller and the GSM module, via the $T_x$ and $R_x$ pins. *RST* pin hard resets the module when pulled to ground. The *DTR* pin wakes the device from periodic inactivity sleep, through $+V_{cc}$ voltage.

*I. NEO-6M GPS Module*

The GPS module is chosen to have minimum power consumption, high localization accuracy and an in-built real-time clock to store date and time information. NEO-6M® GPS module was chosen as it satisfied most of the requirements. The module has a 5Hz position update rate and an EEPROM to save configuration settings with a backup battery. See consumption details from [11].

- Power supply           2.7 V – 3.6 V
- Backup power           1.4 V – 3.6 V, 22 μA
- Power consumption
  *111 mW @ 3.0V (continuous)*
  *33 mW @ 3.0V Power Save Mode (1 Hz)*
  *68 mW @ 1.8V (continuous)*
  *22 mW @ 1.8V Power Save Mode (1 Hz)*

The NEO-6M® module is compatible with UART communication and has relatively low power consumption. Hence, was deemed suitable to be used in the device[13].

*J. Power management module*

The power management module was designed to charge the Li-ion battery and deliver power to the device. The microcontroller operates at 5 V voltage level but, as the battery used in the device was 4.2 V, the power management system contained a DC voltage boost module to boost the battery voltage to 5V. Therefore, a battery charger module TP4056® that contains a 2.1A boost module was used. As shown in Figure V.

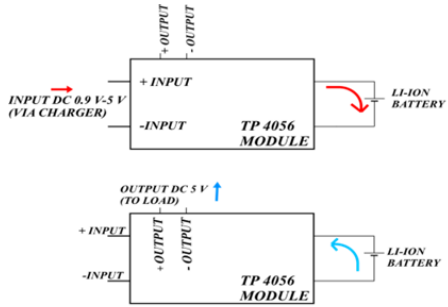

FIGURE V

Power management module schematic. TP4056® functions as a charging module for the Li-ion battery and as a power supply module for the load. During the discharging cycle of the battery, power will flow through the boost converter and provide power to the Microcontroller and to the ancillaries.

*K. Application*

An android based Smartphone app was developed simultaneously to Setup/Configure/Locate/Call/SMS the child device. Figure VI. shows a few screenshots of the application that was developed. The application runs the Route deviation algorithm mentioned in Section IV.

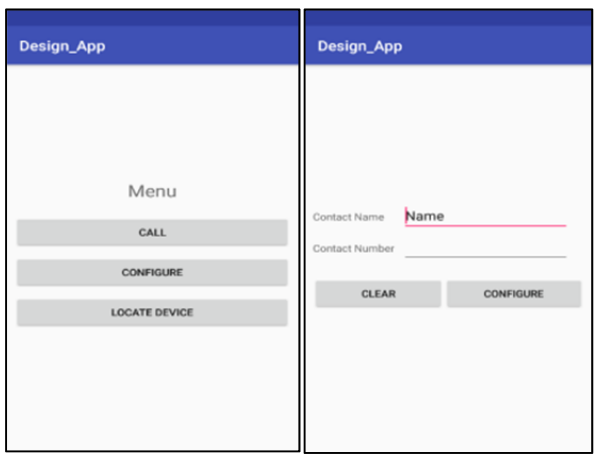

FIGURE VI

Screenshots of the smart device application which runs on an Android platform that allows the parent device to Call/Configure/Locate/track the child device.

## ROUTE DEVIATION DETECTION

In this section, a matrix based method to calculate the probability of the device taking a specified route is presented. Here, an 11 m x 11 m square resolution is used. Increasing the resolution (for example to 1 m x 1 m) will provide more accurate localization but at the cost of high data processing rates. It was decided that an 11 m x 11 m square for localization was sufficient for the device considering inherent percentage errors in GPS data.

The construction method of the probability matrix is given in chronological order in Fig 7-12 and the flow chart representation is given in Fig 14.

As shown in Figure VII the GPS data (Latitude and Longitude) of an 11mx 11m square differs at the fourth decimal place.

| Lat: 7.2545 Lon:80.5970 | Lat: 7.2546 Lon:80.5970 | Lat: 7.2547 Lon:80.5970 |
| --- | --- | --- |
| Lat: 7.2545 Lon:80.5971 | Lat: 7.2546 Lon:80.5971 | Lat: 7.2547 Lon:80.5971 |
| Lat: 7.2545 Lon:80.5972 | Lat: 7.2546 Lon:80.5972 | Lat: 7.2547 Lon:80.5972 |
| Lat: 7.2545 Lon:80.5973 | Lat: 7.2546 Lon:80.5973 | Lat: 7.2547 Lon:80.5973 |

FIGURE VII

The table shows how the GPS location tags differ at the fourth decimal place at a resolution of 11m. This will be used to assign a GPS tags for each grid position.

The grid was laid on a local topogrophical map such that each 11mx11m square on the map contained a location tag. Figure VIII shows the grid to be used in locating the device to a 11mx11m square.

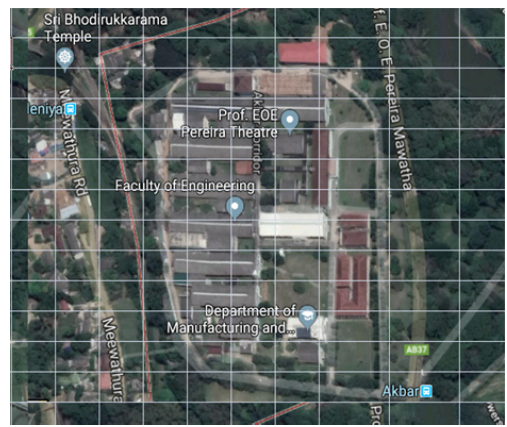

FIGURE VIII

Assignment of the squares according to the GPS data to 11mx11m grid points.

The value of one was assigned to every square identified by its grid location, as shown in Figure IX. This provided a basis for a mathematical model to apply probability and statistics.

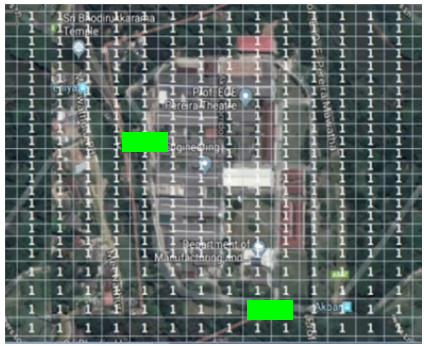

FIGURE IX

Assignment of value one to the squares according to the GPS data to 11mx11m grid points.

For each iteration, if the device is pinged at a square, the value in each square is increased by 1 as shown in Fig X-XII, such that the value in each square is proportional to the probability of the device being pinged at that grid point.

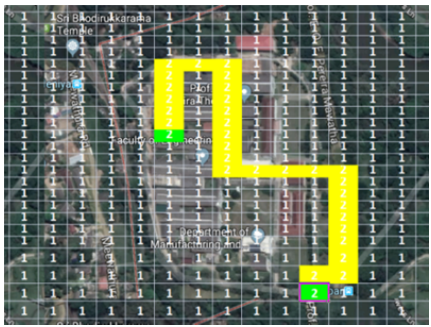

FIGURE X

Grid after one trip is shown above, the squares visited once shown in yellow will have the value of 2. The start and finish positions are shown in green.

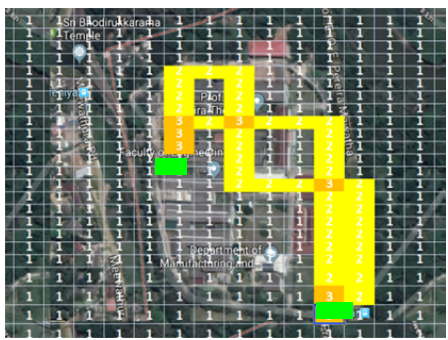

FIGURE XI

The grid after two trips is shown above, the squares visited twice are shown in orange will have the value of 3, the start and finish positions are shown in green.

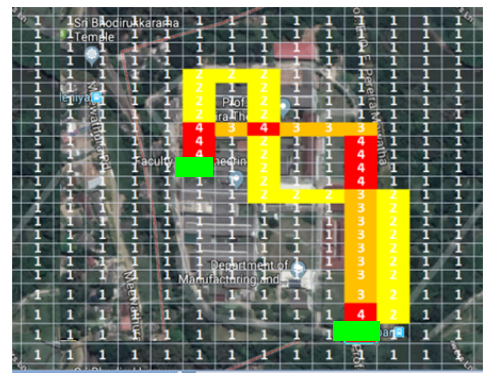

FIGURE XII

The grid after three trips is shown above, the squares visited thrice are shown in red and will have a value of 4, the start and finish positions are shown in green

Probabilities of being at each location are given from the Equation (1) as shown.

$$P_{i,j} = \frac{n_{i,j}}{N} \quad (1)$$

Where $n_{i,j}$ denotes the number of times the device is pinged at the grid point $i,j$ and $N$ is the total number of trips the device takes.

Calculated probabilities are added to the *probability matrix P* as shown in Equation (2). Figure XIII provides a visual representation of matrix *P*. And the flow char of the propeosed algorithm in Figure XIV.

$$P = \begin{pmatrix} \cdot & \cdot & \cdot \\ \cdot & P_{i,j} & \cdot \\ \cdot & \cdot & \cdot \end{pmatrix} \quad (2)$$

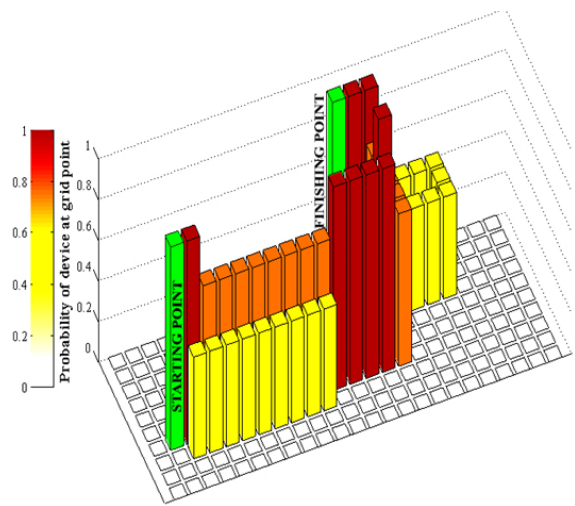

FIGURE XIII

The column chart version of the probability matrix is shown in the figure above. The grid points (squares) visited the most times will have the peaks and the least visited grid points (squares) will have falls

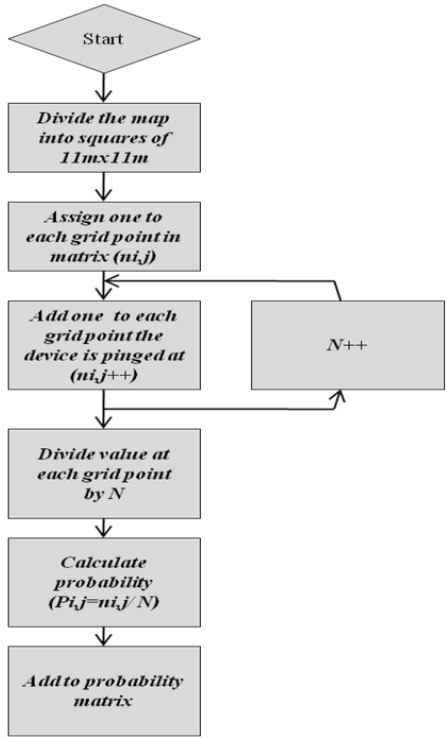

**FIGURE XIV**
The Flowchart representation of the construction of the probability matrix. The figure shows the flow of data in the route deviation algorithm.

As an elementary model, the events of the device getting pinged at different grid points were assumed to be independent. Hence, the total probability of the device taking a certain route was calculated by multiplying probabilities of each of the squares visited, as shown in Equation (3).

$$P' = \prod P_{i,j} \qquad (3)$$

Where $P_{i,j}$ denotes the probability assigned to each grid point from Equation (1) and $P'$ denotes the probability of the device taking a certain route

The Probability of taking a certain route ($P'$) when gone below a preset threshold value, will trigger an alarm in the application on the parent device. The application builds up a database of location data where, a 120 day data retention period will ensure the processing device will not get flooded.

For first three days the device will learn the usual route the child takes. The data is then sent to the application on the parent device where it stores the data in the ROM(Read Only Memory).

**Data privacy** was taken into consideration when designing the device. Location data is not sent to a third party cloud/database where they might be at risk. Rather, the data processing and storage is done on the authorized parent device itself.

## RESULTS

The device was fabricated and tested for accuracy and reliability. Figure XV. shows a photograph of the device in its final merchantable form. The device was tested for performance and battery life. Under normal operating conditions the device had a standby mode power consumption of 50 mW and 115 mW in continuous mode. Battery life depends on the capacity of the battery used. Upon testing it was found that a 1000mAh battery was sufficient for 12 hour continuous usage.

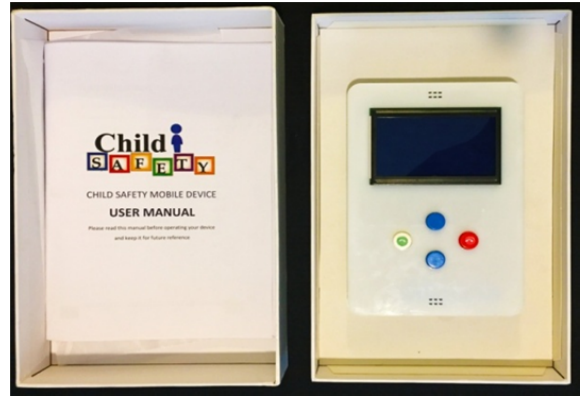

**FIGURE XV**
A photograph of the device with the complete outer packaging and user manual.

Upon Alpha and Beta testing, some problems were identified. This section will discuss some of the major issues that were identified and were promptly rectified. Any future work that will further improve this device will be discussed in Section VI. Further problematic areas were with the hardware operation of the GPS module and the GSM module. Design solutions to avert those problems are mentioned henceforth.

*a) GPS module problems and solutions*

It was identified that the GPS module had accuracy problems which typically resulted in having errors in location data. This was due to the module antenna having difficulty in picking up and maintaining satellite signal. Moreover, the GPS module demanded a current spike of 100 mA at 3.6 V. The stand-alone power supply was incapable of producing such a current impulse, hence causing the device to reboot continuously. To avert the above mentioned problems, the GPS module antenna was re-designed to be placed at the topmost surface of the device, giving it the highest chance of capturing a signal. Moreover, the last known location is saved in the EEPROM in order to avoid getting completely invalid location data. The power supply was backed by an additional capacitor (in parallel with the onboard) to meet the current spike demand.

*b) GSM module problems and solution*

The GSM module utilizes I2C protocol for data transfer. The uploading and downloading data used string type 64-bit words in processor registers. The SRAM of the microcontroller was inadequate to handle such word lengths at the desired baud rates hence, producing erroneous or garbage data. The microcontroller was chosen to have *SRAM >4 KB* to meet the baud rate and word length demand of I2C protocol.

## CONCLUSION

This paper suggested a method for parents to call, locate and track their children using a child-friendly mobile device. After the design, fabrication and testing was done it was identified that the following improvements to the device can further be implemented.

- In the above model, the event of going from one grid point to another is assumed to be independent, but as with most practical situations the probability of a fixed route is more likely. With a fixed route moving from one square to another cannot be considered as an independent event. Thus, the probability should be calculated by getting the dependent probability of each grid point.
- A Method for Route Prediction Based on Hidden Markov Models [14] could be implemented in the future. This will allow the device to pre-estimate the route the device will take.
- The device was suggested to be manufactured as a wearable device for ergonomic convenience. This will require a PCB(Printed Circuit Board) with SMD (Surface Mounted Devices). As the scalability of the product was considered when designing the device, the basic design of circuitry will remain same.
- The accuracy of the GPS module was around 10m outdoor and 25m indoor. These figures can be dramatically reduced by using a GPS module with a higher accuracy figure.